\documentclass[aps,prl,reprint,superscriptaddress, longbibliography]{revtex4-1}
\usepackage{graphicx}
\usepackage{hyperref}

\begin{document}

\title{Explaining the mechanism of random lasing based sensing}

\author{Michele \surname{Gaio}}
\email{michele.gaio@kcl.ac.uk}
\author{Soraya \surname{Caixeiro}}
\affiliation{Department of Physics, King's College London, Strand, London WCR
2LS, United Kingdom.}

\author{Benedetto Marelli}
\altaffiliation{Present address: Department of Civil and Environmental Engineering MIT, 77 Massachusetts
Avenue, Cambridge, MA 02139-4307, USA}
\affiliation{Department of Biomedical Engineering Tufts University 4 Colby Street,
Medford, MA 02155, USA}

\author{Fiorenzo \surname{Omenetto}}
\affiliation{Department of Biomedical Engineering Tufts University 4 Colby Street,
Medford, MA 02155, USA}

\author{Riccardo \surname{Sapienza}}
\affiliation{Department of Physics, King's College London, Strand, London WCR
2LS, United Kingdom.}

\date{\today}

\begin{abstract}

Here we report a random lasing based sensor which shows pH sensitivity exceeding by 2-orders of magnitude that of a conventional fluorescence sensor. We explain the sensing mechanism as related to gain modifications and lasing threshold nonlinearities.  A dispersive diffusive lasing theory matches well the experimental results, and allow us to predict the optimal sensing conditions and a maximal sensitivity as large as 200 times that of an identical fluorescence-based sensor. The simplicity of operation and high sensitivity make it promising for future biosensing applications.

\end{abstract}

\maketitle
Fluorescence based sensing exploiting spontaneous emission
is among the most widespread mechanism for biochemical detection~\cite{Schreml2011,Geng2015}.
Latest developments have focussed on improving the biochemistry of
the fluorescent binder~\cite{Demchenko2009} and on expanding the
monitored functionalities~\cite{Kuimova2009}, as well as on engineering
nanoscale light fields via surface plasmons~\cite{Deng2012}, microcavities~\cite{Vollmer2002},
photonic crystals~\cite{Soboleva2009} or optical resonators~\cite{Akselrod2012}
to enhance light-matter interaction.

Lasing instead, which is based on stimulated emission, has been largely
overlooked as a sensing transducer, mainly because of the complexity
of a conventional lasing architecture. Lasing has the potential to
outperform fluorescence due to the signal amplification inherent to
the lasing process, increased signal-to-noise ratio, narrow emission
line, and non-linear dynamics, as it has been shown for laser-based
interleukin sensing~\cite{Wu2014}, explosives detection~\cite{Rose2005},
and remote identification of hazardous chemicals~\cite{Hokr2014}.
Only recently biocompatible lasing architectures made with vitamin~\cite{Nizamoglu2013}, and
proteins~\cite{Choi2015, Caixeiro2016}
have been fabricated, indicating a path for laser-based biosensing
inside living tissues~\cite{Fan2014a}. 

While conventional lasing requires periodic geometries or carefully
aligned cavities, random lasing (RL) occurs in disordered systems
with optical gain~\cite{Wiersma2008} ranging from semiconductor
powers~\cite{Cao1999,vanderMolen2007} to biomaterials such as human
tissue~\cite{Polson2004}. The lack of an optical cavity gives this
structure resilience against deformation and makes it appealing for
implantation in biological media. Despite the inherit randomness of
RL, emission control has been achieved both spectrally~\cite{gottardo2008,Bachelard2014}
and directionally~\cite{Hisch2013}, and its rich modal properties
have just started to be explored~\cite{Antenucci2015,Leonetti2011}.

Sensing with RL has been limited so far to the detection of changes
of the scattering strength of the matrix by a refractive index~\cite{Polson2004,choi2014}
or temperature~\cite{Wiersma2001} variation. Instead, the potential
of targeted sensing via biochemical interaction at the gain level,
affecting the amplification process, is largely unexplored. 

\begin{figure}[b]
\includegraphics{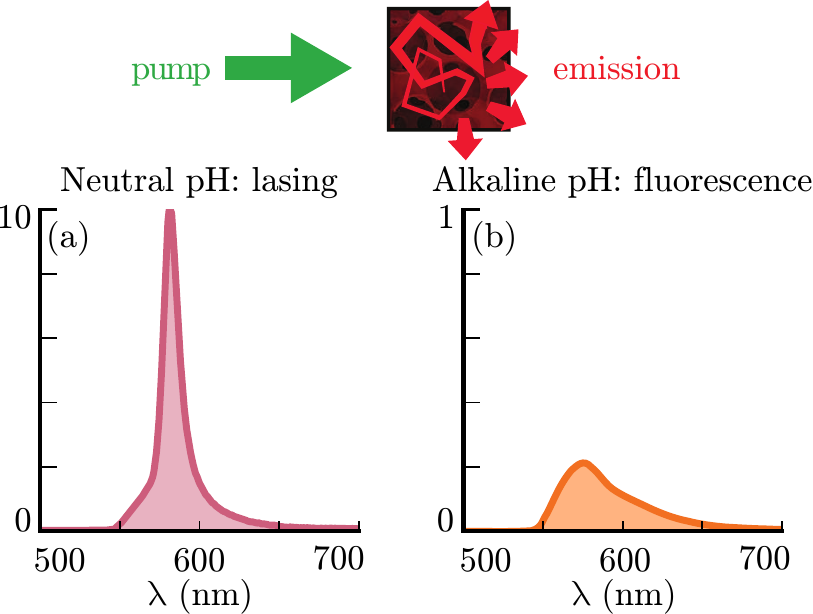} 
\caption{\label{fig:sketch}Random lasing sensing scheme. Light multiple scattering in
the gain medium embedded in a photonic glass leads to amplification
and lasing. This is experimentally visible in the emission spectrum
which shows a narrow band emission (red line, panel (a)). For alkaline
pH the lasing emission is switched off resulting in the broadband
fluorescence emission and lower intensity (orange line, panel (b)). }
\end{figure}

We have recently demonstrated a bio- compatible random laser~\cite{Caixeiro2016},  reacting to pH changes which provides a preliminary sensing proof.
In this letter we explain the mechanism of sensing by random lasing based on gain variation
upon interaction with the biochemical environment: the lasing action
at neutral pH (Fig.~\ref{fig:sketch}(a)), is suppressed at alkaline
pH (Fig.~\ref{fig:sketch}(b)). The experiments are in very good agreement
with the calculations of a dispersive diffusive lasing model without
free parameters. We predict the optimal sensing conditions and we
show that the random lasing sensitivity can be up to 
$200$ times that of fluorescence.

\begin{figure}
\includegraphics{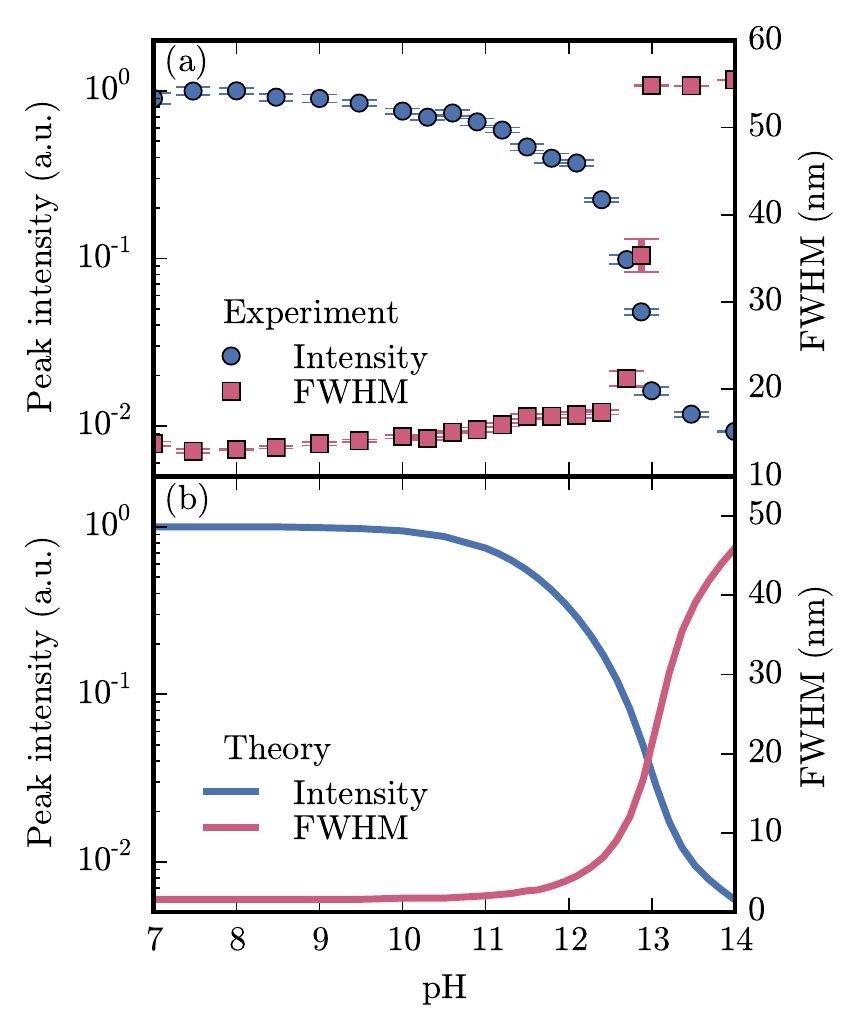}\caption{\label{fig:experiment_and_theory}
Sensing of pH: comparison of experiments and theory. (a) The random lasing system is pumped above the lasing threshold
(P\,=\,840\,$\mu$J/mm$^{2}$) and the emission characteristics as
a function of the pH of the solution are recorded. The lasing is suppressed
at large pH values (pH\,$>13$), corresponding to a strong
decrease of the peak intensity (blue circles) and a sharp increase
of the FWHM of the emission (red square). (b) Theoretical
prediction of the lasing response upon pH variation which shows a
similar behaviour.}
\end{figure}

The random lasing system is fabricated by self-assembly of an inverse
silk photonic glass~\cite{garcia2010} with embedded laser dye (Rhodamine-6G)
as detailed in Ref.~\cite{Caixeiro2016}. The pores have a diameter
of $\sim$1.3~$\mu$m which optimizes the optical scattering (the
measured transport mean free path in water is $\ell_{t}\simeq14~\mu$m).
The sample (thickness $L\simeq100$~$\mu$m)
is excited at a fixed power P\,=\,840~$\mu$J/mm$^{2}$ with 6~ns
pulses of a Nd:YAG Q-switched (wavelength 532~nm, spot diameter$\sim2$~mm)
well above the lasing threshold ($T\simeq80~\mu$J/mm$^{2}$). The
pH of the solution surrounding the laser is controlled by varying
the concentration of NaOH. As shown in Fig.~\ref{fig:experiment_and_theory}(a),
a progressive decrease in peak intensity (blue circles) is observed
for increasing pH: beyond the value pH\,=\,13, the lasing action is
switched off. The peak intensity shows an overall $\sim\,$100-fold
intensity decrease, and the full width at half maximum (FWHM) (red
squares) instead increases smoothly from 14~nm at pH\,=\,$7$, corresponding
to the above-threshold linewidth (shown in Fig.~\ref{fig:sketch}(a)), to pH\,$\simeq13$ where it sharply
reaches 54~nm, which is the FWHM of the fluorescence spectrum (shown in Fig.~\ref{fig:sketch}b)). The
error bars are calculated as the standard deviation of the average of 10 repeated
measurements, each by pumping with a single laser pulse. 

This sensing dynamics can be predicted by a dispersive diffusive lasing
model, built on light diffusion coupled to classical molecular rate
equations and which includes spectral mode competition~\cite{Gaio2015}.
This model has no free parameters, it describes the realistic sample
characteristics given the scattering and gain properties
of the medium either measured or taken from literature~\cite{Caixeiro2016,Holzer2000}, 
while the gain cross-section ($\sigma_{e}$)
is calculated by assuming the same scaling of the absorption ($\sigma_{a}$)~\cite{Holzer2000}.
The resulting theoretical predictions are shown in Fig.~\ref{fig:experiment_and_theory}(b)
and are in very good agreement with the experimental data. As expected
\cite{Gaio2015}, the predicted lasing linewidth is underestimated,
as in the model the narrowing is limited only by the gain saturation. 

Qualitatively, we can understand the lasing switching-off as due to
a reduction of the optical amplification which increases the gain length
$\ell_{g}$ (the distance required for amplification of a factor $e$):
when the critical lasing size $L_{cr}\propto\sqrt{\ell_{t}\ell_{g}}$
required for lasing becomes larger than the sample size, the lasers
switches off. Here the transport mean free path $\ell_{t}$ is unchanged
by pH variations, whereas the gain length $\ell_{g}$ is instead pH
sensitive. More quantitatively, in the approximation of a stationary
and uniform system, the lasing threshold $T$ can be expressed as
\begin{equation}
T\propto[(N\tau_{c}v\sigma_{e}-1)\tau_{r}\Phi\sigma_{a}]^{-1},\label{eq:rl_threshold_formula}
\end{equation}
where $N$ is the molecules density, $v$ the speed of light in the
medium, $\tau_{c}$ is the Thouless time (the typical time it takes
for a photon to escape the disordered medium) which accounts for the
losses at the surface, and the relevant properties of the molecules
providing optical gain are modelled with the stimulated emission cross-section
$\sigma_{e}$, the absorption cross section $\sigma_{a}$ at the pump
wavelength, the radiative lifetime of the excited state $\tau_{r}$,
and the quantum efficiency $\Phi$. The latter two are related via
the non radiative decay rate $\Gamma{}_{nr}=1/\tau_{nr}$, as $\Phi=\Gamma_{r}/(\Gamma_{r}+\Gamma_{nr}).$ 

\begin{figure}
\includegraphics{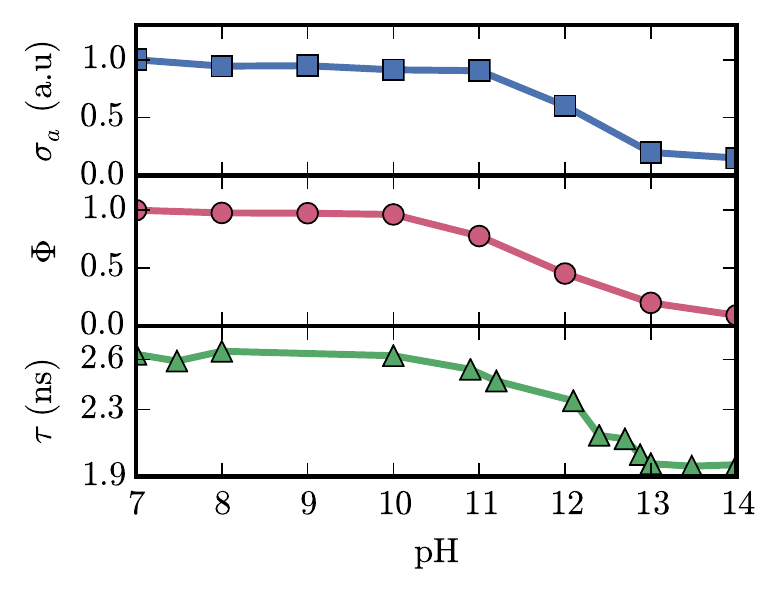}

\caption{\label{fig:dye_properties}Rhodamine-6G properties as a function of pH. The relative
absorption cross-section ($\sigma_{a}$), quantum efficiency ($\Phi$)
and excited state lifetimes ($\tau$) measured as a function of pH.
All quantities decrease for pH values larger than\,$\simeq10$.}
\end{figure}

A change in any of the molecular parameters in Eq.~1 would modify
the lasing threshold and be detectable by the lasing sensor. The dye
fluorescence parameters measured as a function of pH, which are the
input of the lasing model, are shown in Fig.~\ref{fig:dye_properties}. 
The dye properties are unaffected in the pH range 7--10. Starting
at pH\,$\simeq10$, we observe a pronounced decrease of absorption
and lifetime, and from pH~$\simeq 11$ a similar decrease of the
quantum efficiency. The quantum efficiency and lifetime are obtained
from fluorescence studies of the porous and doped silk matrix with
picosecond pulsed excitation ($\lambda=$\,532~nm, 40~MHz): $\Phi$
is obtained as the variation of the fluorescence intensity when recording
the light escaping from the sample with an integrating sphere, and
$\tau$ by fluorescence lifetime spectroscopy by time correlated single
photon counting.  The absorption is measured with a spectrophotometer
and is consistent with lifetime and quantum efficiency as calculated
by the Strickler-Berg relation~\cite{Strickler1962}. No significant
emission spectral shift is observed. It is evident when comparing
Fig.~\ref{fig:dye_properties} and Fig.~\ref{fig:experiment_and_theory}
that the changes in the molecular properties are amplified by the
lasing system and this results in a large intensity variation with
a sharp transition of the lasing emission: this offers opportunity
for efficient sensing.

\begin{figure*}
\includegraphics{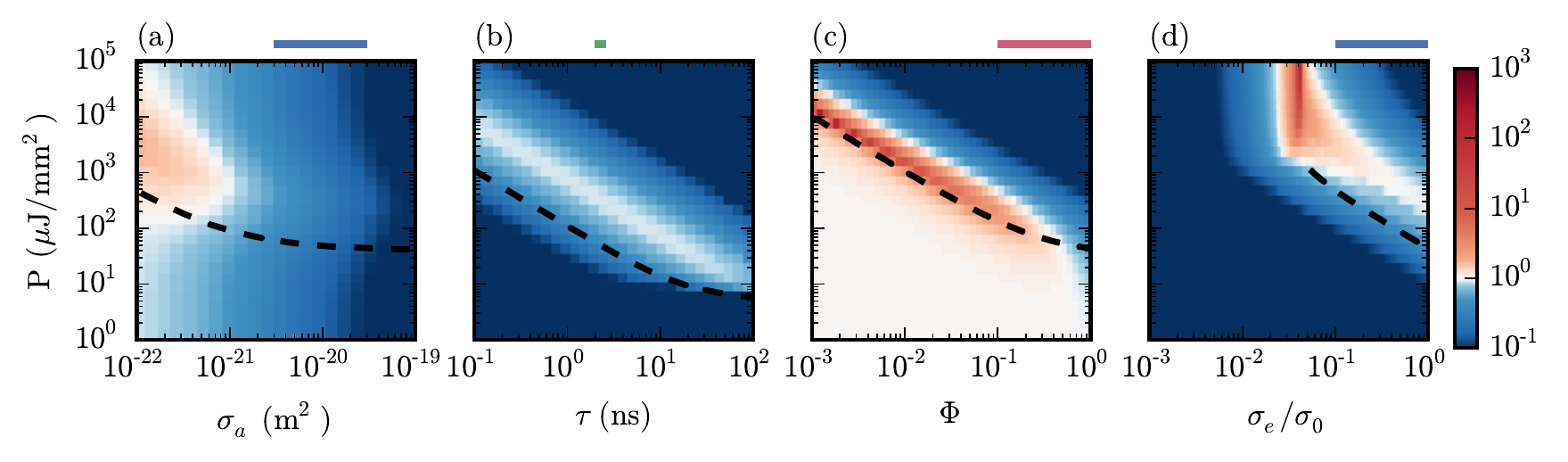}\caption{\label{fig:sensitivity}Sensitivity analysis. The relative sensitivity defined as
$S_{\alpha}=\left|\frac{dI/I}{d\alpha/\alpha}\right|$, for $\alpha=\sigma_{a}$, 
$\tau$, $\Phi$, $\sigma_{e}$, is calculated for the same system
parameters, when varying the value of $\alpha$, and for different
pump intensities. The black dashed lines are the lasing threshold
marking the boundary between the fluorescence and lasing regime. The
blue areas correspond to no-sensitivity ($S_{\alpha}\ll1$), the white
areas correspond to linear sensitivity ($S_{\alpha}=1$), and the
red areas correspond to increased sensitivity ($S_{\alpha}\gg1$).
The highest sensitivities are found around the fluorescence-lasing
transition, with maximum values $S_{\sigma_{a}}=2.2$, $S_{\tau}=0.9$,
$S_{\Phi}=201$, $S_{\sigma_{e}}=186$. The top bars refer to the
measured variation of each parameter as shown in Fig.~\ref{fig:dye_properties}.
}
\end{figure*}

The sensitivity capabilities and limits of RL sensing can be predicted
by calculating the effect of the dye parameters on the lasing threshold.
Although these parameters are typically coupled in real dyes, we consider
them independently to isolate their role. We have chosen as figure
of merit the peak intensity ($I$) relative sensitivity, defined as:
\begin{equation}
S_{\alpha}=\left|\frac{dI/I}{d\alpha/\alpha}\right|,\label{eq:sensitivity}
\end{equation}
where $\alpha$ is the parameter examined. The linear response typical
of the fluorescence regime would give $S_{\alpha}=1$. In Fig.~\ref{fig:sensitivity}
we compute $S_{\alpha}$ for the various molecular parameters ($\alpha=\sigma_{a}$, $\tau$, $\Phi$,
$\sigma_{e}$) at different pump intensities.
The colormap highlights regions with linear response (white) and highly
nonlinear response above $S_{\alpha}=1$ (red). The blue areas correspond
to regions with little or negligible effect on the measured intensity.
The black dashed lines are the calculated lasing threshold, marking
the boundary between the fluorescence and lasing regimes. For all
parameters there are regions with increased sensitivity when compared
to fluorescence ($S_{\alpha}>1$). 

Fig.~\ref{fig:sensitivity} can be understood by considering the
role of the different molecular parameters in the lasing process.
$\sigma_{a}$ describes the pump absorption and therefore the excitation
probability of the fluorophores. This is a typical property exploited
in fluorescence sensing as it induces a variation of the measured
emitted light intensity. In the regime where RL has sizes exceeding the penetration depth of
the pump, a change in pump absorption can be compensated by an increase
of the active volume inside the system, such that the total available
gain is the same,\emph{ i.e.} for large absorption values (right side
of Fig.~\ref{fig:sensitivity}(a)) the RL is insensitive to changes
in $\sigma_{a}$. Instead, for lower absorption values (left side
of Fig.~\ref{fig:sensitivity}(a)), when the pump absorption length
is comparable with the system size, the absorbed intensity and the
emission intensity are linearly related to $\sigma_{a}$ both for
fluorescence and lasing (white areas). Interestingly, around the lasing
threshold (the black line), a twofold increase in the sensitivity $S_{\sigma_{a}}=2.2$
(light red region) is predicted.

A similar behaviour can be observed for $\tau$ which instead describes
the lifetime of the population of the excited state and therefore
is related to the ease of inducing population inversion. As shown
in Fig.~\ref{fig:sensitivity}(b), the recorded intensity is largely
insensitive to a change of $\tau$, both for fluorescence and lasing.
Instead, a lifetime decrease induces a mild shift of the lasing threshold
towards higher pump intensities resulting in a roughly linear sensitivity,
with $S_{\tau}=0.9$ around the lasing threshold. 

The quantum efficiency $\Phi$ is another quantity often exploited
in fluorescence sensing techniques, as it relates directly to the
emitted intensity. The wide linear (white) region below lasing threshold
in Fig.~\ref{fig:sensitivity}(c) is the linear sensitivity of the
fluorescence regime. The lasing emission intensity well above threshold
 is marginally affected by the quantum efficiency,
because non-radiative decay processes are slower than stimulated emission
and therefore they become irrelevant. Instead, around
the lasing threshold the sensitivity peaks, up to $S_{\Phi}=201$,
as shown by the red region. This can be understood as the emission
intensity increases rapidly as stimulated emission (unaffected by
$\Phi$) takes over spontaneous emission (affected by $\Phi$). 

The most direct way of tuning the lasing threshold is by controlling
the gain value, \emph{i.e.} altering $\sigma_{e}$ as shown in Fig.~\ref{fig:sensitivity}(d).
This is a parameter unique to lasing, which has no effect on fluorescence.
As expected, the calculated fluorescence sensitivity is independent
from $\sigma_{e}$ and once again the largest response is found around
the lasing threshold. In this case, a decrease of $\sigma_{e}$ to
roughly 10\% of the original value results in the suppression of the
lasing emission, regardless of the pump intensity. In these conditions,
high sensitivity ($S_{\sigma_{e}}=186$) is reached.

We can now discuss the experimental sensing profile reported in Fig.~\ref{fig:experiment_and_theory}.
The top bars in Fig.~\ref{fig:sensitivity} identify the measured
variation of the parameters. As expected a pH variation affects
all of them, but most notably the gain and quantum efficiency. The
resulting experimental sensitivity extracted from the experimental
data is $S_{\text{pH}}=200\pm50$ at pH\,=\,13, which is larger 
than the theoretical expectation of $S_{\text{pH}}\sim60$. Finally,
in the range pH~$=12\text{--}13$, we estimate a limit of detection (LOD)
of LOD$_{\text{pH}}\simeq0.03$, defined as 3 times the signal
to noise ratio.

Stimulated emission can therefore boost the sensitivity of a fluorescence
sensor as well as provide an additional sensing parameter, \emph{i.e.}
$\sigma_{e}$. These advantages come at the expense of additional
complexity. Lasing requires a nanophotonic architecture to promote
stimulated emission, a disordered medium for RL, and a dye capable
of providing net optical gain. The large $\sim(10\,\mu m)^{3}$ laser
volume implies that the RL sensor is not suitable for sensitivity
at the single molecule level. When compared to  fluorescence
schemes, RL requires a higher excitation intensity, in the $\mu$W
range ($>1~\mu J$ pulse energy) instead of the n$W$ range of conventional
single molecule spectroscopy. While this could be a problem for in
vivo sensing, preliminary results in living cells~\cite{Schubert2015,Humar2015}
show that these power ranges are below the damage threshold of the
biological media. 

In conclusion, we have introduced a RL sensing scheme based on the
lasing threshold shift upon modification of the gain dye parameters.
We have presented a detailed description of the sensing mechanism
and a theoretical model which matches very well the experiments on
pH sensing by silk-based random lasing. We have identified the most
efficient sensing scheme, with a 2-order of magnitude enhancement
with respect to fluorescence. Given the universality of multiple scattering,
its robustness against stress and deformation, and the large availability
of fluorescent and lasing dyes, we foresee possible applications for
bio and chemical sensing in living tissues.

\begin{acknowledgments}
We wish to thank Duong Van Ta and Francisco Fernandes for fruitful
discussions. The research leading to these results has received funding
from the Engineering and Physical Sciences Research Council (EPSRC),
from the European Union, from the Leverhulme Trust and from the Royal
Society. F.G.O. would like to acknowledge the Office of Naval Research
for support of this work (N00014-13-1-0596).
\end{acknowledgments}

\end{document}